\documentclass[twocolumn,tighten]{aastex63}
\usepackage{amssymb, amsmath,framed}

\usepackage{bm}
\expandafter\ifx\csname package@font\endcsname\relax\else
 \expandafter\expandafter
 \expandafter\usepackage
 \expandafter\expandafter
 \expandafter{\csname package@font\endcsname}
\fi
\def\bq{\begin{equation}}
\def\eq{\end{equation}}
\def\bqy{\begin{eqnarray}}
\def\eqy{\end{eqnarray}}

\def\p{\partial}

\def\p{\partial}

\def\R{\mathbb{R}}

 \def\q#1#2{q^{\hspace{1pt}#1}_{\,,\hspace{1pt}#2}}
 \def\a#1#2{a^{\hspace{1pt}#1}_{\,,\hspace{1pt}#2}}
 \def\d#1#2{\delta^{\hspace{1pt}#1}_{\,,\hspace{1pt}#2}}

\begin{document}
\title{\large{Constraints on Aquatic Photosynthesis for Terrestrial Planets Around Other Stars}}

\correspondingauthor{Manasvi Lingam}
\email{mlingam@fit.edu}

\author{Manasvi Lingam}
\affiliation{Department of Aerospace, Physics and Space Sciences, Florida Institute of Technology, Melbourne FL 32901, USA}
\affiliation{Institute for Theory and Computation, Harvard University, Cambridge MA 02138, USA}

\author{Abraham Loeb}
\affiliation{Institute for Theory and Computation, Harvard University, Cambridge MA 02138, USA}

\begin{abstract}
Aquatic photosynthesis plays a major role in carbon fixation and O$_2$ production on Earth. In this Letter, we analyze the prospects for oxygenic photosynthesis in aquatic environments on modern Earth-analogs around F-, G-, K- and M-type stars. Our analysis takes into account the spectral type of the host star, attenuation of light by aquatic organisms, and rates of respiration and photosynthesis. We study the compensation depth ($\mathcal{Z}_\mathrm{CO}$) and the critical depth ($\mathcal{Z}_\mathrm{CR}$), defined respectively as the locations where the net growth rates and vertically integrated net growth rates of photoautotrophs become zero. Our analysis suggests that $\mathcal{Z}_\mathrm{CO}$ declines by more than an order of magnitude as one moves from the habitable zones around Sun-like stars to late-type M-dwarfs, but $\mathcal{Z}_\mathrm{CR}$ decreases by only a modest amount ($\sim 40\%$). For M-dwarf exoplanets, we propose that the photosynthetic red edge may constitute a more robust biosignature of aquatic photosynthesis compared to atmospheric O$_2$. \\
\end{abstract}

\section{Introduction} \label{SecIntro}
The overwhelming majority of carbon fixation and biomass on Earth occurs via oxygenic photosynthesis \citep{Knoll15}. One of the chief reasons behind the proliferation of oxygenic photosynthesis on Earth is that the electron donor (water) was not limited in terms of availability, unlike other variants of photosynthesis \citep{WRF}. The wavelength range for oxygenic photosynthesis on Earth is approximately $350$-$750$ nm \citep{CB11,NMS18}, but most of the light utilized by oxygenic photoautotrophs lies within $\lambda_\mathrm{min} = 400$ nm and $\lambda_\mathrm{max} = 700$ nm, due to which it has been termed photosynthetically active radiation or PAR for short \citep[Chapter 1.2]{Bla14}.\footnote{The upper wavelength of PAR may extend beyond $1000$ nm in principle, but this would entail ``exotic'' multi-photon schemes \citep{WoRa,KST07,Man19} that lie beyond the scope of this work.}

Oceans are known to contribute around $50\%$ of the total primary production on Earth \citep{FB98}. A substantial fraction of terrestrial exoplanets, colloquially referred to as ocean planets, are expected to possess high water inventories, thereby hosting deep oceans and no continents at their surfaces \citep{TI15,ZDR18}. Some of the planets of the well-known TRAPPIST-1 system, for instance, seemingly fall in this category \citep{GDG18,UHD18}. It is therefore essential for studies of planetary habitability to analyze the prospects for aquatic photosynthesis.

However, this field has witnessed comparatively few analyses despite its importance. \citet{WoRa} modeled the global rates of O$_2$ production by Earth-like photoautotrophs at a fixed depth of $10$ m underwater for stars of different spectral types. \citet{KST07} estimated the PAR fluxes at depths of $0.05$ m and $1$ m for various stars, and analyzed the maximum wavelengths at which photosynthesis could operate. However, in order to properly gauge the maximal depths where photoautotrophs may occur, it is necessary to account for biological functions such as respiration and photosynthesis rates. A recent analysis along these lines was undertaken by \citet{RLR18}, but in the specific context of Proxima b. A similar study, ostensibly for the euphotic zone depth, for cool stars is briefly outlined in \citet{Kal19}.

In this Letter, we will examine under what conditions aquatic photosynthesis is feasible and how its essential features are sensitive to the choice of host star. We will incorporate hitherto neglected effects and concepts (e.g., critical depth) and tackle modern Earth-analogs orbiting F-, G-, K- and M-type stars.

\section{Characteristics of aquatic photosynthesis}
We will explore how stellar properties regulate key aspects of aquatic photosynthesis for rocky planets situated in the habitable zones (HZs) of their host stars \citep{KWR93}.

\subsection{Mathematical preliminaries}\label{SSecPrelim}
A rigorous assessment of aquatic photosynthesis requires an in-depth knowledge of biological (e.g., phytoplankton respiration and photosynthesis rates), geological (atmospheric and oceanic composition) and astrophysical (e.g., stellar temperature and flux) parameters. Owing to this complexity, we will hold all factors aside from the stellar properties fixed. The hypothetical planet in question is thus assumed to possess geological and biological attributes akin to Earth. 

The flux incident at the top of Earth's atmosphere is $S_\oplus \approx 1360$ W/m$^2$, and we will suppose that the planet also receives the same amount of stellar flux. Furthermore, for the sake of simplicity, the planet is assumed to be optically thin across the PAR range analogous to modern Earth \citep{Jac99}. Thus, we ignore attenuation of PAR during its passage through the atmosphere. When the star is at the substellar point, the photon flux density at that specific location on the planetary surface (denoted by $\mathcal{N}_\mathrm{max}$) is estimated as
\begin{equation}\label{SpecRadDef}
   \mathcal{N}_\mathrm{max}(\lambda) \approx n_\lambda \left(\frac{R_\star}{d_\star}\right)^2,
\end{equation}
where $R_\star$ is the stellar radius and $d_\star$ is the orbital radius of the Earth-analog, while $n_\lambda$ represents the photon flux density of the star. When the latter is modeled as a black body with an effective stellar temperature of $T$, the photon flux density becomes
\begin{equation}
 n_\lambda = \frac{B_\lambda}{(hc/\lambda)} = \frac{2c}{\lambda^4}\left[\exp\left(\frac{h c}{\lambda k_B T}\right)-1\right]^{-1},
\end{equation}
where $B_\lambda$ is the spectral radiance given by the Planck function. We can express $d_\star$ in terms of the stellar properties by invoking the constraint:
\begin{equation}
    S_\oplus = \frac{L_\star}{4\pi d_\star^2} = \mathrm{const},
\end{equation}
with the stellar luminosity defined as $L_\star = 4\pi \sigma R_\star^2 T^4$. Thus, upon substituting this result in (\ref{SpecRadDef}), we arrive at
\begin{equation}\label{SpecRadMax}
    \mathcal{N}_\mathrm{max}(\lambda) \approx \frac{n_\lambda S_\oplus}{\sigma T^4},
\end{equation}
with the dependence on $R_\star$ being eliminated. However, we note that $\mathcal{N}_\mathrm{max}$ represents the maximum photon flux density because it ignores the effects of clouds and is calculated at the zenith, thus ignoring the rotation of the planet. A more realistic measure of the photon flux density ($\mathcal{N}_\mathrm{avg}$), constituting its temporal average, is
\begin{equation}\label{SpecRadavg}
    \mathcal{N}_\mathrm{avg}(\lambda) \approx \mathcal{N}_\mathrm{max}(\lambda) \cdot f_\mathrm{I} \cdot f_\mathrm{CL},
\end{equation}
where $f_\mathrm{I}$ accounts for the variation in the intensity of light at a given location, and $f_\mathrm{CL}$ embodies the additional attenuation introduced by clouds \citep[Chapter 4.2]{SG06}. For planets that are not tidally locked, $f_\mathrm{I} \approx 1/4$ because the stellar radiation is intercepted across a cross-sectional area of $\pi R^2$ (where $R$ is the planetary radius) but is subsequently distributed over the total surface area of $4\pi R^2$.\footnote{The same result is obtained if the intensity is modeled as a triangular function of time \citep[Chapter 4.2]{SG06}.} However, for tidally locked exoplanets, the radiation must be evenly distributed over the surface area of $2\pi R^2$ because of the permanently dark nightside, which yields $f_\mathrm{I} \approx 1/2$. Determining the stellar ``cutoff'' at which Earth-analogs become tidally locked is difficult because the tidal locking timescale depends on numerous factors such as initial spin period, tidal dissipation factor, and presence/absence of moons \citep{Barn17}.

Next, we turn our attention to the cloud fraction. General circulation models suggest that tidally locked planets at the inner edge of the HZ may manifest relatively high cloud coverage on their dayside; the resulting planetary albedo could become twice that of Earth \citep[Section 3]{YCA13}. Thus, on the one hand, $f_\mathrm{I}$ is elevated for tidally locked exoplanets. On the other, $f_\mathrm{CL}$ is potentially lower due to the greater attenuation from clouds. Hence, we will hold $f_A \equiv f_\mathrm{I} \cdot f_\mathrm{CL}$ fixed as our ensuing results will exhibit a logarithmic dependence on this parameter. We specify $f_A \approx 0.2$ to preserve consistency with Earth's parameters \citep[Chapter 4.3]{SG06}, which transforms (\ref{SpecRadavg}) into $\mathcal{N}_\mathrm{avg}(\lambda) \approx 0.2 \mathcal{N}_\mathrm{max}(\lambda)$. 

The photon flux $\mathcal{F}$ at a depth $z$ below the surface can be determined by employing
\begin{equation}\label{IntRel}
    \mathcal{F}(z) \approx \int_{\lambda_\mathrm{min}}^{\lambda_\mathrm{max}} \mathcal{N}_0(\lambda) \exp\left[-K(\lambda) z\right]\,d\lambda,
\end{equation}
where $\mathcal{N}_0(\lambda)$ is the photon flux density at the surface and is set by either $\mathcal{N}_\mathrm{max}$ or $\mathcal{N}_\mathrm{avg}$ depending on the context. In the above formula, $K(\lambda)$ represents the vertical attenuation coefficient that is further decomposed into $K = K_W + K_C + K_B$, where $K_W$, $K_C$ and $K_B$ denote the partial attenuation coefficients arising from clear water, chemical (both organic and inorganic) compounds, and biota, respectively \citep[Chapter 9.5]{Kirk11}. In actuality, $K_W$, $K_C$ and $K_B$ are complex functions of the wavelength and depth, thereby rendering subsequent calculations difficult to undertake.

We will therefore restrict our scope to encompass two distinct cases henceforth. In both instances, to simplify matters, we set $K_C \rightarrow 0$ and adopt
\begin{equation}
    K_W(\lambda) \approx 1.4 \times 10^{-5}\,\mathrm{m}^{-1}\,\exp\left(\lambda \cdot 1.54 \times 10^7\,\mathrm{m}^{-1}\right),
\end{equation}
across the PAR range because $K_W(\lambda)$ is well approximated by an exponential function; the corresponding data were taken from \citet[Table 3]{PF97}. In the first, we select $\mathcal{N}_0(\lambda) = \mathcal{N}_\mathrm{max}$ and $K_B \rightarrow 0$, which constitutes the most optimal scenario wherein the star is at the zenith and no biological attenuation in present. In the second, we utilize $\mathcal{N}_0(\lambda) = \mathcal{N}_\mathrm{avg}$ and $K_B \approx 0.08$ m$^{-1}$ \citep[Chapter 4.2]{SG06}. This setup is more realistic because the dual affects of temporally averaged stellar flux and biological attenuation in water are incorporated. The two cases will be henceforth be identified by the use of the ``M'' (i.e., maximal) and ``R'' (i.e., realistic) superscripts. 

Although we draw upon the salient characteristics of phytoplankton, this does not imply that the same organisms would necessarily evolve on other worlds; instead, it is merely assumed that their functional attributes are similar. The chief rationale behind employing eukaryotic phytoplankton as a proxy for putative aquatic biota is that they constitute the dominant source of carbon fixation in Earth's present-day oceans \citep{Fal04,Rav09}, and comprise the bedrock of current aquatic ecosystems \citep{Val15}. It is, therefore, worth exploring how a modern Earth-like aquatic biosphere would fare on Earth-analogs around other stars. 

At this stage, a comment regarding the euphotic zone is in order. This zone is typically defined as the depth ($\mathcal{Z}_\mathrm{E}$) at which the intensity is $1\%$ of its surface value \citep[Chapter 6.3]{Kirk11}. For an Earth-analog around a Sun-like star, most of the radiation that penetrates to a depth greater than few meters lies within the PAR range. Hence, utilizing the prior expressions for $K$ yields $\mathcal{Z}_\mathrm{E}^{(M)} \approx 277$ m and $\mathcal{Z}_\mathrm{E}^{(R)} \approx 37$ m. The latter exhibits good agreement with the empirically derived range of $4.3-82.0$ m \citep{LWK07} and the theoretical mean value of $38$ m estimated in \citet[Chapter 4.2]{SG06} for $\mathcal{Z}_\mathrm{E}$. In contrast, for an Earth-analog orbiting a late-type M-dwarf, most of the incident radiation falls within the near-infrared (near-IR), thereby yielding $K \gtrsim 1$ m$^{-1}$ \citep{KLC93}. As per the above definition, we obtain $\mathcal{Z}_\mathrm{E} \lesssim 5$ m, which is consistent with analyses by \citet{RLR18} and \citet{Kal19}.

\begin{figure}
\includegraphics[width=7.5cm]{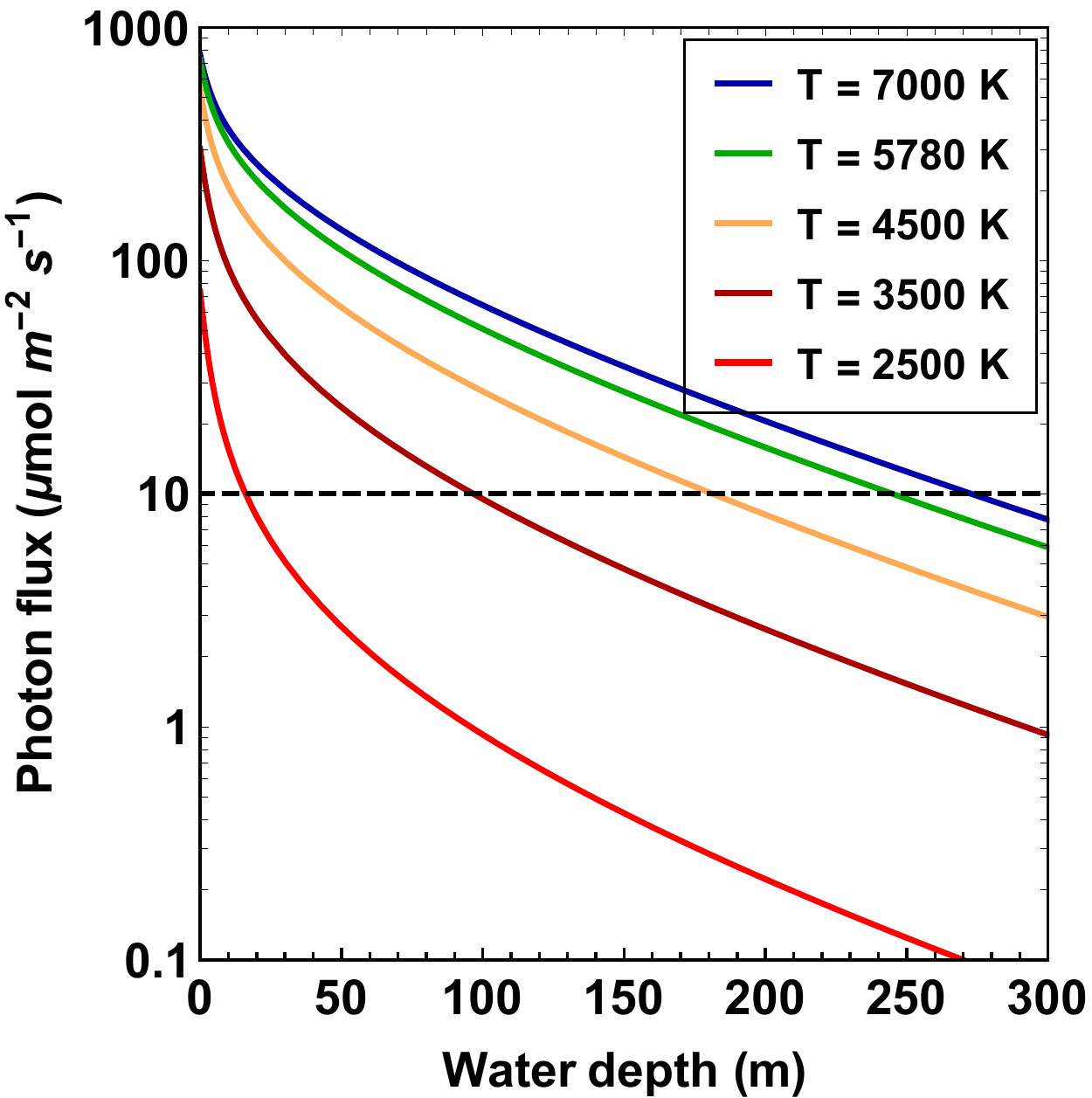} \\
\caption{The maximal photon flux (in $\mu$mol m$^{-2}$ s$^{-1}$) is shown as a function of the depth for Earth-analogs around FGKM stars; the curves correspond to different stellar temperatures. The horizontal dashed line yields the compensation depth, where the rates of photosynthesis and respiration balance each other.}
\label{FigCompDepthO}
\end{figure}

\begin{figure}
\includegraphics[width=7.5cm]{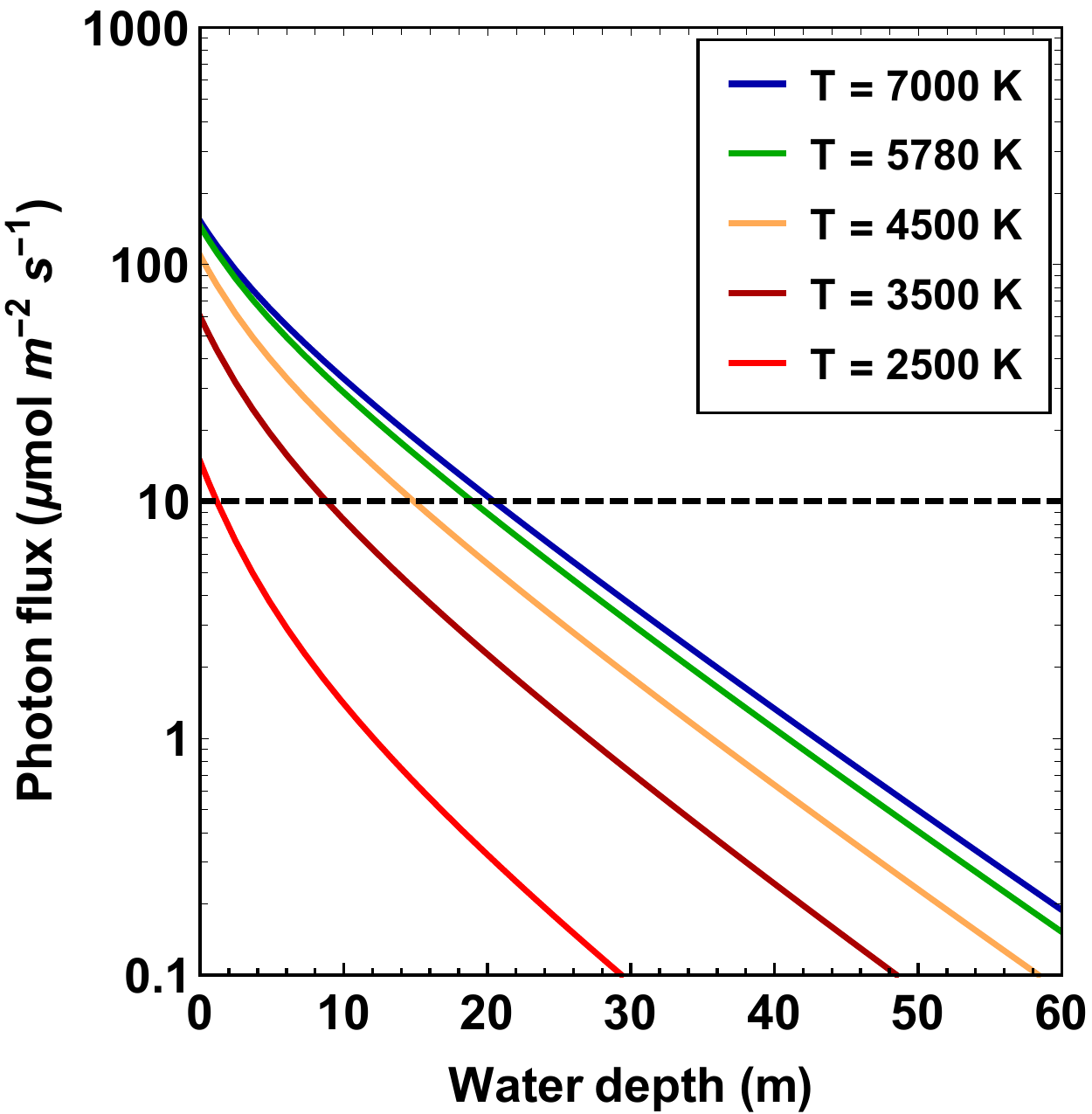} \\
\caption{The photon flux (in $\mu$mol m$^{-2}$ s$^{-1}$) is shown as a function of the depth for Earth-analogs around FGKM stars; the curves correspond to different stellar temperatures. The horizontal dashed line yields the compensation depth, where the rates of photosynthesis and respiration balance each other. In this model, time-averaged photon flux and aquatic biological attenuation are incorporated.}
\label{FigCompDepthR}
\end{figure}

\subsection{Critical depth and compensation depth}
Although the euphotic zone depth has some intrinsic value, it does not yield information concerning the maximal depths where photoautotrophs can exist. In order to estimate these quantities, we will analyze the \emph{compensation} and \emph{critical} depths, both of which were elucidated in the seminal work by \citet{Sver53}. 

The compensation depth ($\mathcal{Z}_\mathrm{CO}$) is defined at the location where the rate of photosynthesis is sufficient to balance the respiration rate. At greater depths, respiration will dominate over photosynthesis, thus inhibiting the growth of the photosynthetic community. The photon flux at which this critical balance occurs is the compensation flux ($\mathcal{F}_C$). By supposing that the oxygenic photoautotrophs are akin to phytoplankton on Earth, we specify $\mathcal{F}_C \sim 10$ $\mu$mol m$^{-2}$ s$^{-1}$ \citep{RD10,RLR18}; varying $\mathcal{F}_C$ by a factor of $\sim 2$ \citep{NS91,SDY02} exerts a minor influence on subsequent results via (\ref{IntRel}). Thus, by calculating the location where $\mathcal{F}(z) = \mathcal{F}_C$ is attained, one can duly determine the approximate location of the compensation depth. 

In Fig. \ref{FigCompDepthO}, the maximal photon flux has been plotted as a function of the depth for Earth-analogs orbiting different stars. By making use of (\ref{IntRel}) and the threshold $\mathcal{F}_C$, we find that $\mathcal{Z}_\mathrm{CO}^{(M)} \approx 244$ m for a Sun-like star ($T = 5780$ K),\footnote{This result seems compatible with the detection of microalgae at depths of $285$ m on Earth \citep[Chapter 3.1]{Val15}.} and $\mathcal{Z}_\mathrm{CO}^{(M)} \approx 16$ m for a late-type M-dwarf analogous to TRAPPIST-1 ($T = 2500$ K). These numbers are in reasonable agreement with the corresponding values of $185$ m and $10$ m calculated for Earth and Proxima b, respectively \citep[Fig. 7]{RLR18}. 

Next, we turn our attention to the case where time-averaged photon flux and biological attenuation are included. The ensuing results are depicted in Fig. \ref{FigCompDepthR}. For a Sun-like star, we estimate $\mathcal{Z}_\mathrm{CO}^{(R)} \approx 20$ m whereas we find $\mathcal{Z}_\mathrm{CO}^{(R)} \approx 1$ m for a late-type M-dwarf with $T = 2500$ K. The former value is consistent with empirical estimates of $\sim 20$-$30$ m for the compensation depth in multiple environments \citep{SJF42,SDY02,Mid19}. For both the maximal and realistic cases, we find that the compensation depth is reduced by more than an order of magnitude as one moves from Sun-like stars to the coolest M-dwarfs.

To undertake a similar calculation for anoxygenic photoautotrophs, two changes must be implemented. First, the longest wavelength suitable for photosynthesis must be extended to $\lambda_\mathrm{max} \approx 1000$ nm \citep[Chapter 1.2]{Bla14}. Second, the photon flux at the compensation point is specified to be $\mathcal{F}_C \sim 1$ $\mu$mol m$^{-2}$ s$^{-1}$ based on the \emph{Chlorobium} species extracted from lakes and fjords in Vestfold Hills, Antarctica \citep{BB88}.\footnote{As per empirical data and theoretical constraints, $\mathcal{F}_C \sim 0.01$ $\mu$mol m$^{-2}$ s$^{-1}$ is compatible with anoxygenic photoautotrophs \citep{RKB00,MGK05}, but the prior conservative limit is adopted for comparison against \citet{RLR18}.} We can neglect the first factor without much loss of generality because water is strongly absorbing in the near-IR \citep{KLC93}. 

Upon utilizing this value of $\mathcal{F}_C$, we obtain $\mathcal{Z}_\mathrm{CO}^{(M)} \approx 500$ m and $\mathcal{Z}_\mathrm{CO}^{(M)} \approx 95$ m for a Sun-like star and late-type M-dwarf ($T = 2500$ K), respectively. In comparison, a similar analysis by \citet[Fig. 8]{RLR18} yielded compensation depths of approximately $400$ m and $60$ m for Earth and Proxima b. For the realistic scenario described previously, we find $\mathcal{Z}_\mathrm{CO}^{(R)} \approx 40$ m and $\mathcal{Z}_\mathrm{CO}^{(R)} \approx 12$ m for the solar analog and the late-type M-dwarf.

\begin{figure}
\includegraphics[width=7.5cm]{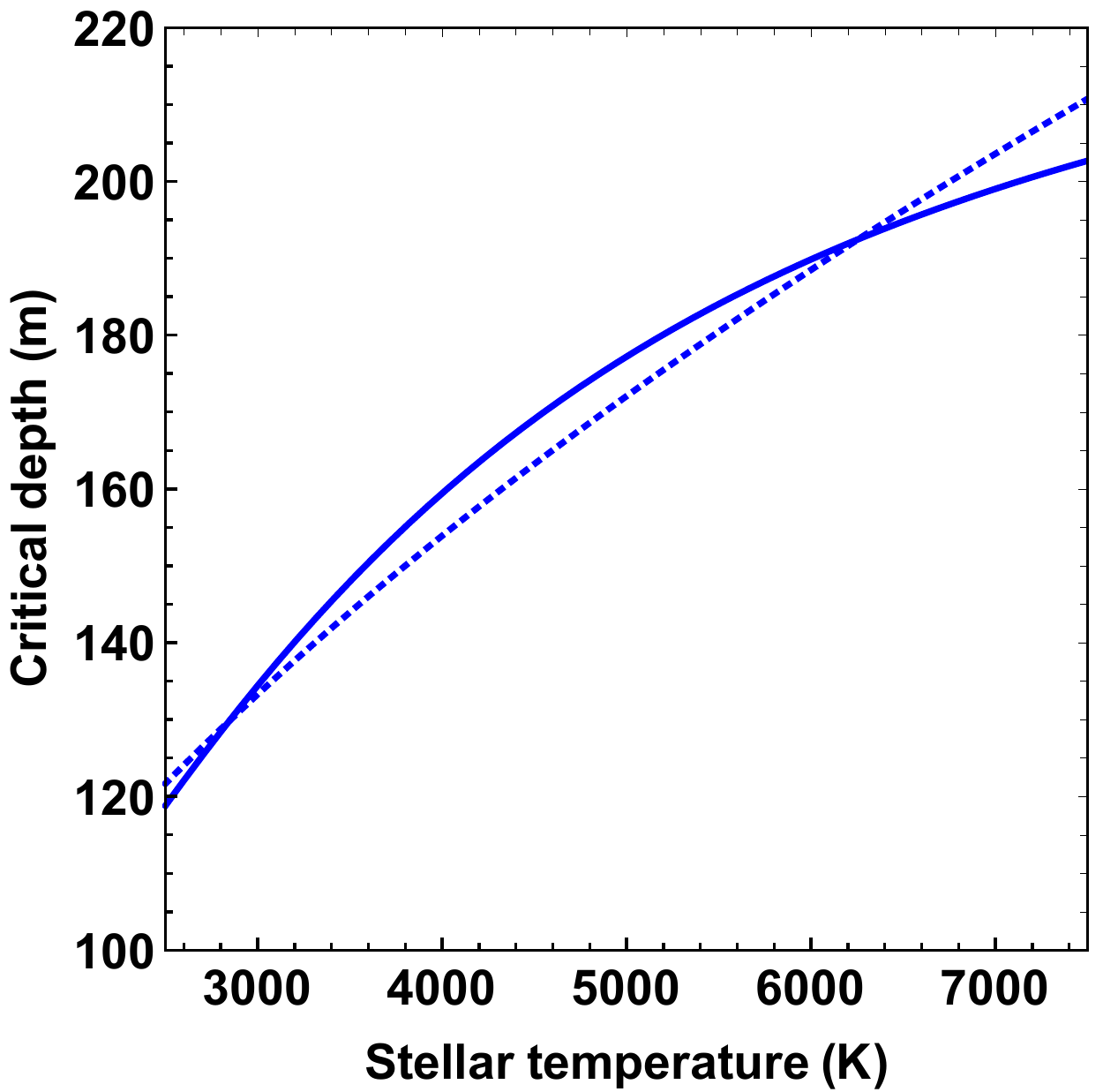} \\
\caption{Critical depth (in m), the location at which the vertically integrated net growth rate becomes zero, as a function of stellar temperature (in K). Effects of time-averaged photon flux and aquatic biological attenuation are incorporated. The unbroken line corresponds to (\ref{CritD}), whereas the dotted line depicts the power-law approximation given by $\mathcal{Z}_\mathrm{CR} \approx 185\,\mathrm{m}\, \sqrt{T/T_\odot}$.}
\label{CritDepth}
\end{figure}

Now, we turn our attention to gauging the critical depth ($\mathcal{Z}_\mathrm{CR}$), namely, the thickness of the aquatic layer where the vertically integrated photosynthetic growth rate exceeds the total loss rate due to respiration and other factors \citep[Chapter 3]{ML06}. Thus, communities circulating in this layer are theoretically capable of survival and growth.\footnote{Although the critical depth hypothesis constitutes a vital concept in biological oceanography, some of its underlying postulates and consequent predictions have been challenged \citep{Behr10}.} The estimation of $\mathcal{Z}_\mathrm{CR}$ is not straightforward because a number of divergent (albeit cognate) formulae exist: see \citet[equation 4.3.5]{SG06}, \citet[equation 11.1]{Kirk11}, \citet[equation 3.08]{ML06} and \citet[equation 2.27]{Mid19}. The expression provided in the last two references is equivalent to the classic result derived by \citet{Sver53}, and equals
\begin{equation}\label{CdepthEq}
    \frac{1 - \exp\left(-K \mathcal{Z}_\mathrm{CR}\right)}{K \mathcal{Z}_\mathrm{CR}} = \frac{\Gamma_R}{\Gamma_P},
\end{equation}
where $\Gamma_R$ and $\Gamma_P$ denote the rates of respiration and maximal photosynthesis, respectively. This can be further simplified to yield $K \mathcal{Z}_\mathrm{CR} \approx \Gamma_P/\Gamma_R$ \citep[equation 9.7]{FR07} because $K \mathcal{Z}_\mathrm{CR} \gg 1$ is valid.

As the above formula was obtained under the assumption of $K = \mathrm{const}$, it is necessary to recalculate $\mathcal{Z}_\mathrm{CR}$ for $K(\lambda)$. After implementing the same procedure \citep[Chapter 3]{ML06}, we arrive at
\begin{equation}\label{CritD}
  \mathcal{Z}_\mathrm{CR} \approx \left(\frac{\Gamma_R}{\Gamma_P}\right)^{-1} \frac{\int_{\lambda_\mathrm{min}}^{\lambda_\mathrm{max}} \left[\mathcal{N}_0(\lambda)/K(\lambda)\right]\,d\lambda}{\int_{\lambda_\mathrm{min}}^{\lambda_\mathrm{max}} \mathcal{N}_0(\lambda)\,d\lambda}
\end{equation}
We specify $\Gamma_R/\Gamma_P \approx 3.36 \times 10^{-2}$ for putative biota \citep[Chapter 4.3]{SG06} and adopt the parameters from the ``R'' case introduced earlier to facilitate comparison with prior studies that accounted for time-averaged photon flux and biological attenuation.

The resultant critical depth is plotted in Fig. \ref{CritDepth}. The heuristic formula $\mathcal{Z}_\mathrm{CR} \approx 185\,\mathrm{m}\, \sqrt{T/T_\odot}$ displays excellent agreement ($< 5\%$) with the actual results. From (\ref{CritD}), we obtain $\mathcal{Z}_\mathrm{CR}^{(R)} \approx 187$ m for the solar analog. This result compares favorably with the estimate of $170 \pm 30$ m for Earth's oceans \citep{SDY02,SG06} and $177$ m for Lake Windermere, England \citep[Chapter 11.1]{Kirk11}. For a late-type M-dwarf with $T = 2500$ K, we arrive at $\mathcal{Z}_\mathrm{CR}^{(R)} \approx 119$ m. 

Thus, as evinced by Fig. \ref{CritDepth}, the critical depth is relatively insensitive to the stellar temperature. This trend probably arises because the bulk of productivity occurs at shallow depths, where the rates of photosynthesis are much higher due to their near-linear dependence on $\mathcal{F}(z)$ \citep{ML06,SG06}; as $\mathcal{Z}_\mathrm{CR}$ entails vertical integration, most of the contribution to net growth is from the upper layers. In principle, therefore, the extent of the zone wherein the integrated net growth of phytoplankton-like organisms is feasible remains roughly constant across Earth-analogs orbiting different stars. Moreover, as $\mathcal{Z}_\mathrm{CR}$ governs the initiation of phytoplankton blooms \citep{FR07}, \emph{ceteris paribus}, analogous phenomena may have a similar likelihood of occurrence on these worlds.

\section{Discussion and Conclusions}
We estimated the compensation and critical depths for Earth-analogs around various stars. We determined that the former decreases by more than an order of magnitude as one moves from solar analogs to the smallest stars (i.e., late-type M-dwarfs); in contrast, the critical depth varies by merely $\sim 40\%$ across the same range. 

Our work has several implications for life detection. Due to the lower compensation depth associated with late-type M-dwarfs, the rates of carbon fixation could be correspondingly lower, which is consistent with prior analyses of this subject \citep{WoRa,RLR18}. Earth-analogs around these stars are expected to have lower likelihoods of building up oxygenated atmospheres because of diminished O$_2$ production rates \citep{LCPH,LM19}. This would, in turn, give rise to ``false negatives'' insofar as life detection through atmospheric oxygen is concerned. Inability to accumulate atmospheric O$_2$ might also prove to be detrimental for the origin of complex multicellularity due to metabolic constraints \citep{CGZM,LL19}.

Even if biologically oxygenated atmospheres are suppressed on M-dwarf exoplanets, the biomass density near the surface is nevertheless potentially comparable to that of Earth's oceans. In fact, due to the combined action of a slower rotation rate (induced by tidal locking) and stronger tidal forces, nutrient upwelling could increase on these worlds, thereby conceivably elevating the biomass density \citep{LL18,OJA}. Hence, for sufficiently high coverage and density of oxygenic photoautotrophs, the photosynthetic red edge (PRE) may facilitate the detection of life. In the absence of cloud cover and $50\%$ surface coverage by oceanic cyanobacteria, \citet[Table 1]{OK19} estimated that the reflected flux would increase by $10\%$ at the PRE. Hence, for such tidally locked planets, the reflected flux at the PRE ought to vary between $0\%$ and $\sim 10\%$ over an orbital period, thus possibly rendering this biofeature amenable to detection.

The expression of a surface signature from anoxygenic photosynthesis or non-photosynthetic organisms may have occurred on the Archean Earth prior to the evolution of oxgenic photosynthesis and detectable levels of its atmospheric signature \citep{SKP18}. The PRE, however, is widely considered unlikely to have been detectable prior to the emergence of vegetation on land \citep{LRP14}. Based on modeling by \citet{OK19}, our work suggests that the PRE might be discernible even in the absence of a detectable O$_2$ biosignature.

There are several caveats that merit reiteration. Perhaps most importantly, we assumed that the functional traits of putative photoautotrophs were akin to eukaryotic phytoplankton. The spectral diversity and flexibility of cyanobacteria analogs, especially their proven capacity for utilizing chlorophylls d and f at far-red and near-IR wavelengths \citep{NMS18,SKP18}, might render them increasingly important for Earth-analogs orbiting cool stars. As the predominant marine cyanobacteria species can grow at photon fluxes that are $\sim 10$ times smaller than the compensation flux considered herein \citep[Chapter 3.2.1]{CKT05}, our results must be revised upward by a factor of $\lesssim 10$ in accordance with Figs. \ref{FigCompDepthO} and \ref{FigCompDepthR}.

We also neglected deviations from the black body spectrum and the deleterious effects of stellar flares. However, with regards to the latter, a combination of screening compounds, strong absorption by water at ultraviolet wavelengths, and physiological adaptations (e.g., DNA repair) may collectively ensure that organisms are protected several meters underwater \citep{CM98,LL19}. Last, but not least, we have not addressed the crucial issue of nutrient limitation in this Letter. Ocean planets, for instance, have been predicted to possess limited biospheres due to low rates of phosphate supply from weathering \citep{WP13,LiMa}.

Despite these caveats, our model retains sufficient complexity (without sacrificing simplicity), consequently enabling us to make concrete and testable predictions. In particular, if future spectroscopic and photometric observations of M-dwarf exoplanets detect no evidence of biotic O$_2$ in the atmosphere and find evidence for the PRE, this would lend credence to the notion that these worlds might host an unusual combination of fairly dense but shallow aquatic biospheres.

\acknowledgments
This research was supported in part by the Breakthrough Prize Foundation, Harvard University's Faculty of Arts and Sciences, and the Institute for Theory and Computation (ITC) at Harvard University.


\begin{thebibliography}{}
\expandafter\ifx\csname natexlab\endcsname\relax\def\natexlab#1{#1}\fi
\providecommand{\url}[1]{\href{#1}{#1}}
\providecommand{\dodoi}[1]{doi:~\href{http://doi.org/#1}{\nolinkurl{#1}}}
\providecommand{\doeprint}[1]{\href{http://ascl.net/#1}{\nolinkurl{http://ascl.net/#1}}}
\providecommand{\doarXiv}[1]{\href{https://arxiv.org/abs/#1}{\nolinkurl{https://arxiv.org/abs/#1}}}

\bibitem[{{Barnes}(2017)}]{Barn17}
{Barnes}, R. 2017, Celest. Mech. Dyn. Astron., 129, 509,
  \dodoi{10.1007/s10569-017-9783-7}

\bibitem[{{Behrenfeld}(2010)}]{Behr10}
{Behrenfeld}, M.~J. 2010, Ecology, 91, 977, \dodoi{10.1890/09-1207.1}

\bibitem[{{Blankenship}(2014)}]{Bla14}
{Blankenship}, R.~E. 2014, {Molecular Mechanisms of Photosynthesis}, 2nd edn.
  (Wiley-Blackwell)

\bibitem[{{Burke} \& {Burton}(1988)}]{BB88}
{Burke}, C.~M., \& {Burton}, H.~R. 1988, Hydrobiologia, 165, 13,
  \dodoi{10.1007/BF00025570}

\bibitem[{{Canfield} {et~al.}(2005){Canfield}, {Kristensen}, \&
  {Thamdrup}}]{CKT05}
{Canfield}, D., {Kristensen}, E., \& {Thamdrup}, B. 2005, {Aquatic
  Geomicrobiology}, Advances in Marine Biology No.~48 (Elsevier)

\bibitem[{{Catling} {et~al.}(2005){Catling}, {Glein}, {Zahnle}, \&
  {McKay}}]{CGZM}
{Catling}, D.~C., {Glein}, C.~R., {Zahnle}, K.~J., \& {McKay}, C.~P. 2005,
  Astrobiology, 5, 415, \dodoi{10.1089/ast.2005.5.415}

\bibitem[{{Chen} \& {Blankenship}(2011)}]{CB11}
{Chen}, M., \& {Blankenship}, R.~E. 2011, Trends Plant Sci., 16, 427,
  \dodoi{10.1016/j.tplants.2011.03.011}

\bibitem[{{Cleaves} \& {Miller}(1998)}]{CM98}
{Cleaves}, H.~J., \& {Miller}, S.~L. 1998, Proc. Natl. Acad. Sci. USA, 95,
  7260, \dodoi{10.1073/pnas.95.13.7260}

\bibitem[{{Falkowski} {et~al.}(2004){Falkowski}, {Katz}, {Knoll}, {Quigg},
  {Raven}, {Schofield}, \& {Taylor}}]{Fal04}
{Falkowski}, P.~G., {Katz}, M.~E., {Knoll}, A.~H., {et~al.} 2004, Science, 305,
  354, \dodoi{10.1126/science.1095964}

\bibitem[{{Falkowski} \& {Raven}(2007)}]{FR07}
{Falkowski}, P.~G., \& {Raven}, J.~A. 2007, {Aquatic photosynthesis}, 2nd edn.
  (Princeton University Press)

\bibitem[{{Field} {et~al.}(1998){Field}, {Behrenfeld}, {Randerson}, \&
  {Falkowski}}]{FB98}
{Field}, C.~B., {Behrenfeld}, M.~J., {Randerson}, J.~T., \& {Falkowski}, P.
  1998, Science, 281, 237, \dodoi{10.1126/science.281.5374.237}

\bibitem[{{Grimm} {et~al.}(2018){Grimm}, {Demory}, {Gillon}, {Dorn}, {Agol},
  {Burdanov}, {Delrez}, {Sestovic}, {Triaud}, {Turbet}, {Bolmont}, {Caldas},
  {de Wit}, {Jehin}, {Leconte}, {Raymond}, {Van Grootel}, {Burgasser}, {Carey},
  {Fabrycky}, {Heng}, {Hernandez}, {Ingalls}, {Lederer}, {Selsis}, \&
  {Queloz}}]{GDG18}
{Grimm}, S.~L., {Demory}, B.-O., {Gillon}, M., {et~al.} 2018, Astron.
  Astrophys., 613, A68, \dodoi{10.1051/0004-6361/201732233}

\bibitem[{{Jacob}(1999)}]{Jac99}
{Jacob}, D.~J. 1999, {Introduction to Atmospheric Chemistry} (Princeton
  University Press)

\bibitem[{{Kaltenegger}(2019)}]{Kal19}
{Kaltenegger}, L. 2019, in AAS/Division for Extreme Solar Systems Abstracts,
  Vol.~51, AAS/Division for Extreme Solar Systems Abstracts, 502.05

\bibitem[{{Kasting} {et~al.}(1993){Kasting}, {Whitmire}, \& {Reynolds}}]{KWR93}
{Kasting}, J.~F., {Whitmire}, D.~P., \& {Reynolds}, R.~T. 1993, Icarus, 101,
  108, \dodoi{10.1006/icar.1993.1010}

\bibitem[{{Kiang} {et~al.}(2007){Kiang}, {Segura}, {Tinetti}, {Govindjee},
  {Blankenship}, {Cohen}, {Siefert}, {Crisp}, \& {Meadows}}]{KST07}
{Kiang}, N.~Y., {Segura}, A., {Tinetti}, G., {et~al.} 2007, Astrobiology, 7,
  252, \dodoi{10.1089/ast.2006.0108}

\bibitem[{Kirk(2011)}]{Kirk11}
Kirk, J. T.~O. 2011, {Light and Photosynthesis in Aquatic Ecosystems}, 3rd edn.
  (Cambridge University Press)

\bibitem[{{Knoll}(2015)}]{Knoll15}
{Knoll}, A.~H. 2015, {Life on a Young Planet: The First Three Billion Years of
  Evolution on Earth}, Princeton Science Library (Princeton University Press)

\bibitem[{{Kou} {et~al.}(1993){Kou}, {Labrie}, \& {Chylek}}]{KLC93}
{Kou}, L., {Labrie}, D., \& {Chylek}, P. 1993, Appl. Opt., 32, 3531,
  \dodoi{10.1364/AO.32.003531}

\bibitem[{{Lee} {et~al.}(2007){Lee}, {Weidemann}, {Kindle}, {Arnone}, {Carder},
  \& {Davis}}]{LWK07}
{Lee}, Z., {Weidemann}, A., {Kindle}, J., {et~al.} 2007, J. Geophys. Res.
  Oceans, 112, C03009, \dodoi{10.1029/2006JC003802}

\bibitem[{{Lehmer} {et~al.}(2018){Lehmer}, {Catling}, {Parenteau}, \&
  {Hoehler}}]{LCPH}
{Lehmer}, O.~R., {Catling}, D.~C., {Parenteau}, M.~N., \& {Hoehler}, T.~M.
  2018, Astrophys. J., 859, 171, \dodoi{10.3847/1538-4357/aac104}

\bibitem[{{Lingam} \& {Loeb}(2018)}]{LL18}
{Lingam}, M., \& {Loeb}, A. 2018, Astrobiology, 18, 967,
  \dodoi{10.1089/ast.2017.1718}

\bibitem[{{Lingam} \& {Loeb}(2019{\natexlab{a}})}]{Man19}
---. 2019{\natexlab{a}}, Astrophys. J., 883, 143,
  \dodoi{10.3847/1538-4357/ab3f35}

\bibitem[{{Lingam} \& {Loeb}(2019{\natexlab{b}})}]{LM19}
---. 2019{\natexlab{b}}, Mon. Not. R. Astron. Soc., 485, 5924,
  \dodoi{10.1093/mnras/stz847}

\bibitem[{{Lingam} \& {Loeb}(2019{\natexlab{c}})}]{LL19}
---. 2019{\natexlab{c}}, Rev. Mod. Phys., 91, 021002,
  \dodoi{10.1103/RevModPhys.91.021002}

\bibitem[{{Lingam} \& {Loeb}(2019{\natexlab{d}})}]{LiMa}
---. 2019{\natexlab{d}}, Astron. J., 157, 25, \dodoi{10.3847/1538-3881/aaf420}

\bibitem[{{Lyons} {et~al.}(2014){Lyons}, {Reinhard}, \& {Planavsky}}]{LRP14}
{Lyons}, T.~W., {Reinhard}, C.~T., \& {Planavsky}, N.~J. 2014, Nature, 506,
  307, \dodoi{10.1038/nature13068}

\bibitem[{{Mann} \& {Lazier}(2006)}]{ML06}
{Mann}, K.~H., \& {Lazier}, J. R.~N. 2006, Dynamics of Marine Ecosystems:
  Biological-Physical Interactions in the Oceans, 3rd edn. (Blackwell
  Publishing)

\bibitem[{{Manske} {et~al.}(2005){Manske}, {Glaeser}, {Kuypers}, \&
  {Overmann}}]{MGK05}
{Manske}, A.~K., {Glaeser}, J., {Kuypers}, M. M.~M., \& {Overmann}, J. 2005,
  Appl. Environ. Microbiol., 71, 8049, \dodoi{10.1128/AEM.71.12.8049-8060.2005}

\bibitem[{{Middelburg}(2019)}]{Mid19}
{Middelburg}, J.~J. 2019, {Marine Carbon Biogeochemistry: A Primer for Earth
  System Scientists}, SpringerBriefs in Earth System Sciences (Springer),
  \dodoi{10.1007/978-3-030-10822-9}

\bibitem[{{Nelson} \& {Smith}(1991)}]{NS91}
{Nelson}, D.~M., \& {Smith}, W.~O. 1991, Limnol. Oceanogr., 36, 1650,
  \dodoi{10.4319/lo.1991.36.8.1650}

\bibitem[{{N{\"u}rnberg} {et~al.}(2018){N{\"u}rnberg}, {Morton},
  {Santabarbara}, {Telfer}, {Joliot}, {Antonaru}, {Ruban}, {Cardona}, {Krausz},
  {Boussac}, {Fantuzzi}, \& {Rutherford}}]{NMS18}
{N{\"u}rnberg}, D.~J., {Morton}, J., {Santabarbara}, S., {et~al.} 2018,
  Science, 360, 1210, \dodoi{10.1126/science.aar8313}

\bibitem[{{Olson} {et~al.}(2019){Olson}, {Jansen}, \& {Abbot}}]{OJA}
{Olson}, S.~L., {Jansen}, M., \& {Abbot}, D.~S. 2019, Astrophys. J.,
  arXiv:1909.02928.
\newblock \doarXiv{1909.02928}

\bibitem[{{O'Malley-James} \& {Kaltenegger}(2019)}]{OK19}
{O'Malley-James}, J.~T., \& {Kaltenegger}, L. 2019, Astrophys. J. Lett., 879,
  L20, \dodoi{10.3847/2041-8213/ab2769}

\bibitem[{{Pope} \& {Fry}(1997)}]{PF97}
{Pope}, R.~M., \& {Fry}, E.~S. 1997, Appl. Opt., 36, 8710,
  \dodoi{10.1364/AO.36.008710}

\bibitem[{{Raven}(2009)}]{Rav09}
{Raven}, J.~A. 2009, Aquat. Microb. Ecol., 56, 177, \dodoi{10.3354/ame01315}

\bibitem[{{Raven} {et~al.}(2000){Raven}, {K{\"u}bler}, \& {Beardall}}]{RKB00}
{Raven}, J.~A., {K{\"u}bler}, J.~E., \& {Beardall}, J. 2000, J. Mar. Biol.
  Assoc. UK, 80, 1, \dodoi{10.1017/s0025315499001526}

\bibitem[{{Regaudie-De-Gioux} \& {Duarte}(2010)}]{RD10}
{Regaudie-De-Gioux}, A., \& {Duarte}, C.~M. 2010, Global Biogeochem. Cy., 24,
  GB4013, \dodoi{10.1029/2009GB003639}

\bibitem[{{Ritchie} {et~al.}(2018){Ritchie}, {Larkum}, \& {Ribas}}]{RLR18}
{Ritchie}, R.~J., {Larkum}, A. W.~D., \& {Ribas}, I. 2018, Int. J. Astrobiol.,
  17, 147, \dodoi{10.1017/S1473550417000167}

\bibitem[{{Sarmiento} \& {Gruber}(2006)}]{SG06}
{Sarmiento}, J.~L., \& {Gruber}, N. 2006, {Ocean Biogeochemical Dynamics}
  (Princeton University Press)

\bibitem[{{Schwieterman} {et~al.}(2018){Schwieterman}, {Kiang}, {Parenteau},
  {Harman}, {DasSarma}, {Fisher}, {Arney}, {Hartnett}, {Reinhard}, {Olson},
  {Meadows}, {Cockell}, {Walker}, {Grenfell}, {Hegde}, {Rugheimer}, {Hu}, \&
  {Lyons}}]{SKP18}
{Schwieterman}, E.~W., {Kiang}, N.~Y., {Parenteau}, M.~N., {et~al.} 2018,
  Astrobiology, 18, 663, \dodoi{10.1089/ast.2017.1729}

\bibitem[{{Siegel} {et~al.}(2002){Siegel}, {Doney}, \& {Yoder}}]{SDY02}
{Siegel}, D.~A., {Doney}, S.~C., \& {Yoder}, J.~A. 2002, Science, 296, 730,
  \dodoi{10.1126/science.1069174}

\bibitem[{{Sverdrup}(1953)}]{Sver53}
{Sverdrup}, H.~U. 1953, J. Cons. Perm. Int. Explor. Mer., 18, 287,
  \dodoi{10.1093/icesjms/18.3.287}

\bibitem[{{Sverdrup} {et~al.}(1942){Sverdrup}, {Johnson}, \& {Fleming}}]{SJF42}
{Sverdrup}, H.~U., {Johnson}, M.~W., \& {Fleming}, R.~H. 1942, {The Oceans:
  Their Physics, Chemistry, and General Biology} (Prentice-Hall)

\bibitem[{{Tian} \& {Ida}(2015)}]{TI15}
{Tian}, F., \& {Ida}, S. 2015, Nat. Geosci., 8, 177, \dodoi{10.1038/ngeo2372}

\bibitem[{{Unterborn} {et~al.}(2018){Unterborn}, {Hinkel}, \& {Desch}}]{UHD18}
{Unterborn}, C.~T., {Hinkel}, N.~R., \& {Desch}, S.~J. 2018, Res. Notes AAS, 2,
  116, \dodoi{10.3847/2515-5172/aacf43}

\bibitem[{{Valiela}(2015)}]{Val15}
{Valiela}, I. 2015, {Marine Ecological Processes}, 3rd edn. (Springer),
  \dodoi{10.1007/978-0-387-79070-1}

\bibitem[{{Ward} {et~al.}(2019){Ward}, {Rasmussen}, \& {Fischer}}]{WRF}
{Ward}, L.~M., {Rasmussen}, B., \& {Fischer}, W.~W. 2019, J. Geophys. Res.
  Biogeosci., 124, 211, \dodoi{10.1029/2018JG004679}

\bibitem[{{Wolstencroft} \& {Raven}(2002)}]{WoRa}
{Wolstencroft}, R.~D., \& {Raven}, J.~A. 2002, Icarus, 157, 535,
  \dodoi{10.1006/icar.2002.6854}

\bibitem[{{Wordsworth} \& {Pierrehumbert}(2013)}]{WP13}
{Wordsworth}, R.~D., \& {Pierrehumbert}, R.~T. 2013, Astrophys. J., 778, 154,
  \dodoi{10.1088/0004-637X/778/2/154}

\bibitem[{{Yang} {et~al.}(2013){Yang}, {Cowan}, \& {Abbot}}]{YCA13}
{Yang}, J., {Cowan}, N.~B., \& {Abbot}, D.~S. 2013, Astrophys. J. Lett., 771,
  L45, \dodoi{10.1088/2041-8205/771/2/L45}

\bibitem[{{Zain} {et~al.}(2018){Zain}, {de El{\'\i}a}, {Ronco}, \&
  {Guilera}}]{ZDR18}
{Zain}, P.~S., {de El{\'\i}a}, G.~C., {Ronco}, M.~P., \& {Guilera}, O.~M. 2018,
  Astron. Astrophys., 609, A76, \dodoi{10.1051/0004-6361/201730848}

\end{thebibliography}

\end{document}